\documentstyle[preprint,prb,aps]{revtex}
\begin{document}
      %\draft command makes pacs numbers print
\draft
\title{Electromagnetic waves in a Josephson junction in a 
thin film}
\author{R.G.~Mints}
\address{School of Physics and Astronomy,\\ Raymond
and Beverly Sacler Faculty of Exact Sciences, \\Tel Aviv
University, \\ Tel Aviv 69978, Israel}
\author{I.B.~Snapiro}
\address{Physics Department,\\ Technion--Israel
Institute of Technology,\\ Haifa 32000, Israel}
\maketitle
\begin{abstract}
      %insert abstract here
We consider a one-dimensional Josephson junction in a 
superconducting film with the thickness that is much less 
than the London penetration depth. We treat an 
electromagnetic wave propagating along this tunnel contact. 
We show that the electrodynamics of a Josephson junction in 
a thin film is nonlocal if the wave length is less than the 
Pearl penetration depth. We find the integro-differential 
equation determining the phase difference between the two
superconductors forming the tunnel contact. We use this 
equation to calculate the dispersion relation for an 
electromagnetic wave propagating along the Josephson 
junction. We find that the frequency of this wave is 
proportional to the square root of the wave vector if the 
wave length is less than the Pearl penetration depth.
\end{abstract}
      %insert suggested PACS numbers in braces on next line

\narrowtext
\section{Introduction}
The electromagnetic properties of tunnel Josephson 
junctions are a subject of intensive studies over the past 
three decades \cite{barone}. A considerable attention is, 
in particular, attracted to the investigation of the 
SIS--type Josephson contacts. In this case the tunnel 
junction is formed by a thin layer of an insulator. This 
dielectric layer between two superconducting plates can be 
treated as a transmission line or a parallel plate 
resonator when the electromagnetic properties are concerned.
It follows from this approach that an electromagnetic wave 
with a specific dispersion relation may propagate along the 
SIS--type Josephson junction \cite{swihart}. 

The existence of this Swihart electromagnetic wave results, 
in particular, in the self-induced resonances, usually 
referred to as Fiske steps \cite{fiske}. These specific 
resonances are observed as peaks in the current-voltage 
curves of the tunnel Josephson junctions. The study of Fiske
steps is one of the methods to treat the electromagnetic 
properties of the SIS--type Josephson contacts. Recently, 
it was successfully applied to investigate the 
electromagnetic properties of the grain boundaries in 
YBa$_2$Cu$_3$O$_{7-\delta}$ high-temperature superconductors
\cite{winkler}.

The self-induced Fiske resonances are observed in the 
presence of an external magnetic field, $H_e$. In this case 
the phase difference between the two superconductors 
forming the tunnel contact, $\varphi$, is a sum of two 
terms, {\it i.e.}, $\varphi=\varphi_H+\varphi_V$. The first 
term, $\varphi_H$, describes the increase of $\varphi$ 
along the Josephson junction ($y$-axis). The value of 
$\varphi_H$ is proportional to the magnetic field $H_e$ and 
is given by the formula 
$\varphi_H=-4\pi\lambda H_ey/\Phi_0$, where 
$\lambda$ is the London penetration depth and $\Phi_0$ is 
the flux quantum. The second term, $\varphi_V$, represents 
the increase of $\varphi$ in time. The value of $\varphi_V$ 
is proportional to the voltage, $V$, applied to the contact,
and is given by the formula $2eVt/\hbar$. As a result the 
tunnel current density
\begin{equation} 
j=j_c\sin\varphi=j_c\sin\Bigl({2eV\over\hbar}\,t-
{4\pi\lambda H_e\over\Phi_0}\,y\Bigr)
\label{ei0}
\end{equation}
is described by a wave propagating along the Josephson 
junction \cite{ekk}. The amplitude of this wave is given by 
the Josephson critical current density, $j_c$.

The self-induced Fiske resonances arise when the frequency 
of the tunnel current density wave, $2eV/\hbar$, becomes 
equal to the frequency of the Swihart electromagnetic wave 
\cite{ekk,kulik}. Thus, the dependence of the frequency, 
$\omega$, on the wave vector, $k$, for an electromagnetic 
wave propagating along the Josephson junction determines 
the positions of the peaks in the current-voltage curve of 
a tunnel contact.

Usually, the dispersion relation $\omega (k)$ is determined 
by the sin-Gordon equation that leads to \cite{barone}
\begin{equation}
\omega=\omega_j\,\sqrt{1+k^2\lambda_J^2},
\label{ei1}
\end{equation}
where 
\begin{equation}
\omega_j=\sqrt{2ej_c\over\hbar C}
\label{ei2}
\end{equation}
is the Josephson frequency, C is the specific capacitance
of the tunnel junction, and    
\begin{equation}
\lambda_j=\sqrt{c\Phi_0\over 16\pi^2\lambda j_c},
\label{ei3}
\end{equation}
is the Josephson penetration depth. 

The Swihart electromagnetic wave corresponds to the limiting
case $1\ll k\lambda_J$. It follows than from Eq.~(\ref{ei1})
that $\omega\approx\omega_J\lambda_J k$, {\it i.e.}, this 
wave is propagating along the Josephson junction with a 
constant velocity
\begin{equation} 
c_s=\omega_J\lambda_J={c\over\sqrt{8\pi\lambda C}}.
\label{ei4}
\end{equation}

The linear dispersion relation $\omega=c_sk$ results in an
equidistant set of peaks in the current-voltage curves of 
the SIS-type Josephson tunnel junctions \cite{ekk,kulik}.

We can determine the phase difference $\varphi (y,t)$ in
the mainframe of the local Josephson electrodynamics, 
{\it i.e.}, by the sin-Gordon equation, as long as 
$k\lambda\ll 1$ \cite{barone}. It means, in particular, 
that for an electromagnetic wave propagating along the 
Josephson junction with $\lambda\ll\lambda_J$ the dispersion 
relation $\omega=c_sk$ is valid in the region
$\lambda_J^{-1}\ll k\ll\lambda^{-1}$. 

Let us now discuss the general case, {\it i.e.}, the case 
when the restrictions on the wave vector $k$ are given 
by the inequalities $kd_0\ll 1$ and $k\xi\ll 1$, where 
$d_0$ is the thickness of the insulating barrier and $\xi$ 
is the coherence length. We treat here an SIS-type 
Josephson junction formed by two superconducting plates. 
In this case the space distribution of $\varphi$ is 
one-dimensional and the relation between the phase 
difference $\varphi (y,t)$ and 
the magnetic field in the superconductors is nonlocal if 
$1\ll k\lambda$. As a result the function $\varphi (y,t)$ 
is determined by an integro-differential equation 
\cite{gurevich,wephysica}, {\it i.e.}, the electrodynamics 
of a Josephson junction is nonlocal as far as the region 
of wave vectors $1\ll k\lambda$ is concerned. 
Using this equation it was shown \cite{weprb,weprl} that 
the dispersion relation for an electromagnetic wave with 
$1\ll k\lambda$ takes the form
\begin{equation}
\omega=c_s\,\sqrt{k\over\lambda}.
\label{ei5}
\end{equation}

The phase velocity of this electromagnetic wave is inversely
proportional to the square root of the wave vector $k$. It 
results in a non-equidistant set of the values of the voltage 
$V$ corresponding to the self-induced resonances 
in the current-voltage curve.\footnote{We present the theory
of the self-induced Fiske resonances in the mainframe of the 
nonlocal Josephson electrodynamics elsewhere.}  

This effect is most pronounced when considering an SIS-type 
Josephson junction in a thin superconducting film with the 
thickness $d\ll\lambda$. In this case the space scale of 
magnetic field variation in the superconductors forming the 
contact is given by the Pearl penetration depth \cite{pearl}
\begin{equation}
\lambda_{\rm eff}={\lambda^2\over d}\gg\lambda.
\label{ei6}
\end{equation} 

Thus, the electrodynamics of a Josephson junction in a thin 
film is nonlocal if $1\ll k\lambda_{\rm eff}$. This region 
of wave vectors is much wider than the one given by the 
inequality $1\ll k\lambda$. 

In this paper we consider an infinite one-dimensional 
Josephson tunnel junction in a superconducting film with 
the thickness $d\ll\lambda$. We show that the 
electrodynamics of a Josephson contact is nonlocal if the 
space scale of variation of $\varphi$ is less than the Pearl 
penetration depth, {\it i.e.}, if $1\ll k\lambda_{\rm eff}$. 
We derive the integro-differential equation determining the 
phase difference $\varphi (y,t)$. We use this equation to 
calculate the dispersion relation $\omega (k)$ for an 
electromagnetic wave propagating along an SIS-type Josephson 
junction. 

The paper is organized in the following way. In Sec. II, we
consider the electrodynamics of a long one-dimensional 
Josephson junction in a superconducting film with the 
thickness that is much less than the London penetration 
depth. We treat the general case of an arbitrary relation 
between the Josephson penetration depth and the effective 
Pearl penetration depth. We derive the integro-differential 
equation determining the phase difference distribution 
along the Josephson junction. In Sec. III, we apply this 
equation to calculate the dispersion relation for an 
electromagnetic wave propagating along the Josephson 
junction. In Sec. IV, we summarize the overall conclusions.

\section{Basic equations}
Let us consider a thin superconducting film ({\em xy}-plane)
with an SIS-type Josephson junction along the {\em y}-axis
as it is shown in Fig.~\ref{f1}. 
We treat here the case when $\lambda\gg\xi$, in which the 
London equations govern the fields and currents inside the 
superconductor. Thus, outside the tunnel contact the 
relation between the current density ${\bf j}$ and the 
magnetic field ${\bf h}$ is given by \cite{tinkham}
\begin{equation}
{\bf h}+{4\pi\lambda^2\over c}\,{\rm rot}\,{\bf j}=0.
\label{be1}
\end{equation}

Introducing the vector potential ${\bf A}$ and combining 
Eq.~(\ref{be1}) with the equation
\begin{equation}
{\bf h}={\rm rot}\,{\bf A}
\label{be2}
\end{equation}
we express the current density {\bf j} in the form 
\begin{equation}
{\bf j}={c\over 4\pi\lambda^2}\,({\bf S}-{\bf A}),
\label{be3}
\end{equation}
where outside the tunnel junction the vector field ${\bf S}$ 
is given by the formula
\begin{equation}
{\bf S}={\Phi_0\over 2\pi}\,\nabla\theta
\label{be4}
\end{equation}
and $\theta$ is the phase of the order parameter.

The quantities ${\bf j}$, ${\bf A}$ and ${\bf S}$ are nearly 
independent on the {\em z} coordinate in the limiting case 
of a thin film, {\it i.e.}, for $d\ll\lambda$. Therefore, in 
order to find the fields and currents we replace the 
superconducting film with the thickness $d\ll\lambda$ by an 
infinitely thin current-carrying sheet in the plane $z=0$. 
The current density ${\bf j}$ in this plane is determined 
then by the averaging of Eq.~(\ref{be3}) over the thickness 
$d$\cite{pearl,degennes}, which results in
\begin{equation}
{\bf j}={c\over 4\pi\lambda_{\rm eff}}\,({\bf S}-{\bf A})\,
\delta (z).
\label{be5}
\end{equation}

Let us now choose the London gauge, {\it i.e.}, let us 
assume that ${\rm div}\,{\bf A}=0$. Then, substituting 
Eq.~(\ref{be5}) into the Maxwell equation
\begin{equation}
{\rm rot}\,{\bf h}={4\pi\over c}\,{\bf j}
\label{be6}
\end{equation}
we find the equation describing the vector potential 
${\bf A}$ in the form 
\begin{equation}
-\Delta {\bf A}+\lambda_{\rm eff}^{-1}{\bf A}\,
\delta (z)=\lambda_{\rm eff}^{-1}{\bf S}\,
\delta (z).
\label{be7}
\end{equation}

The vector field ${\bf S}$ is related to the phase 
difference
\begin{equation}
\varphi (y)=\theta (+0,y)-\theta (-0,y).
\label{be8}
\end{equation}
This relation is given by the equation  
\begin{equation}
{\rm rot}\,{\bf S}={\Phi_0\over 2\pi}\,\varphi'(y)\,
\delta (x)\,\hat{\bf z}, 
\label{be9}
\end{equation}
following from Eq.~(\ref{be4}) and taking into account the 
singularity of the function $\theta (x,y)$ at $x=0$. The 
vector $\hat{\bf z}$ is here for the unit vector along the 
{\em z}-axis. 

Applying the continuity equation ${\rm div}\,{\bf j}=0$ to 
Eq.~(\ref{be5}) we find that ${\rm div}\,{\bf S}=0$. Thus, 
we can present the vector field ${\bf S}$ as a curl of a 
certain vector field ${\bf F}$, namely, 
\begin{equation}
{\bf S(\mbox{\boldmath $\rho$})}={\rm rot}\,{\bf F},
\label{be10}
\end{equation}
where $\mbox{\boldmath $\rho$}=(x,y)$ and
\begin{equation}
{\bf F}=F(\mbox{\boldmath $\rho$})\,\hat{\bf z}.
\label{be11}
\end{equation}
Substituting Eq.~(\ref{be11}) into Eq.~(\ref{be9}) we find 
the equation describing the function 
$F(\mbox{\boldmath $\rho$})$ in the form 
\begin{equation}
\Delta F=-{\Phi_0\over 2\pi}\,\varphi{'}(y)\,\delta (x).
\label{be12}
\end{equation}

Note that the scalar function $F(\mbox{\boldmath $\rho$})$ 
determines both components of the vector field 
${\bf S}(\mbox{\boldmath $\rho$})$ reducing by this way the 
complexity of the problem.

The current density across the Josephson junction $j_x(0,y)$ 
is a sum of two terms, namely, the tunnel and the 
displacement current densities 
\begin{equation}
j_x(0,y)=\bigl[j_c\,\sin\varphi +{\hbar C\over 2e}\,
{\partial^2\varphi\over\partial t^2}\Bigr]d\,\delta (z).
\label{be13}
\end{equation}
At the same time it follows from Eq.~(\ref{be3}) that the 
current density $j_x(0,y)$ can be written as
\begin{equation}
j_x(0,y)={c\over 4\pi\lambda_{\rm eff}}\,
[S_x(0,y)-A_x(0,y,0)]\,\delta (z).
\label{be14}
\end{equation}
Equating the expressions for the quantity $j_x(0,y)$ 
given by Eqs.~(\ref{be13}) and (\ref{be14}) we find that
\begin{equation}
j_c\,\sin\varphi +{\hbar C\over 2e}\,
{\partial^2\varphi\over\partial t^2}=
{c\over 4\pi\lambda^2}\,[S_x(0,y)-A_x(0,y,0)]
\label{be15}
\end{equation} 

Thus, to derive the closed form of the equation describing 
the phase difference $\varphi (y,t)$ it is necessary to find
the functional relation between 
\begin{equation}
\Delta_x(y)=S_x(0,y)-A_x(0,y,0) 
\label{be15a}
\end{equation}
and $\varphi (y,t)$. We use here Fourier transformation in 
order to do it and defining the Fourier transforms for 
${\bf A}({\bf r})$ and {\bf S}(\mbox{\boldmath $\rho$}) as
\begin{equation}
{\bf A}({\bf r})=\int{\bf A}_{{\bf q},p}\,
\exp (i\mbox{\boldmath $q\rho$}+ipz)\,
{d^2{\bf q}dp\over (2\pi)^3}
\label{be16}
\end{equation}
and
\begin{equation}
{\bf S}(\mbox{\boldmath $\rho$})=\int{\bf S}_{\bf q}\,
\exp (i\mbox{\boldmath $q\rho$})\,
{d^2{\bf q}\over (2\pi)^2}.
\label{be17}
\end{equation}

Using ${\bf A}_{{\bf q},p}$ and ${\bf S}_{\bf q}$ we can 
present the value of $\Delta_x$ by the integral 
\begin{equation}
\Delta_x={1\over 4\pi^2}\,\int\limits_0^\infty
q\,dq\int\limits_{-\pi}^\pi d\vartheta\,
(S^x_{\bf q}-A^x_{\bf q})\,\exp (iqy\sin\vartheta),
\label{be18}
\end{equation}
where $\vartheta$ is the polar angle in the
$(q_x,q_y)$-plane and
\begin{equation}
{\bf A}_{\bf q}=\int\limits_{-\infty}^{\infty}
{\bf A}_{{\bf q},p}\,{dp\over 2\pi}.
\label{be19}
\end{equation}

The next step is to apply Fourier transformation to 
Eq.~(\ref{be7}), which results in 
\begin{equation}
(q^2+p^2)\,{\bf A}_{{\bf q},p}+
\lambda_{\rm eff}^{-1}\,{\bf A}_{\bf q}=
\lambda_{\rm eff}^{-1}\,{\bf S}_{\bf q}.
\label{be20}
\end{equation}
It follows from Eq.~(\ref{be20}) that the relation between 
${\bf A}_{\bf q}$ and ${\bf S}_{\bf q}$ has the form
\begin{equation}
{\bf A}_{\bf q}={{\bf S}_{\bf q}\over 
1+2q\lambda_{\rm eff}}
\label{be21}
\end{equation}
and thus
\begin{equation}
S^x_{\bf q}-A^x_{\bf q}={2q\lambda_{\rm eff}\over 
1+2q\lambda_{\rm eff}}\,S^x_{\bf q}.
\label{be22}
\end{equation}

To calculate the Fourier transform $S^x_{\bf q}$ we take 
the derivative of Eq.~(\ref{be12}) with respect to $y$ and
substitute $S_x$ instead of $\partial F/\partial y$.
As a result it comes out that
\begin{equation}
\Delta S_x=-{\Phi_0\over 2\pi}\,\varphi{''}(y)\,\delta (x).
\label{be23}
\end{equation}
It follows from Eq.~(\ref{be23}) that the Fourier 
transform $S^x_{\bf q}$ is given by the formula
\begin{equation}
S^x_{\bf q}={\Phi_0\over
2\pi q^2}\,\int\limits_{-\infty}^\infty 
dy\,\varphi{''}(y)\exp (-iqy\sin\vartheta)
\label{be24}
\end{equation}

Combining now Eqs.~(\ref{be24}), (\ref{be22}), (\ref{be18}), 
and (\ref{be15}) we find the integro-differential equation 
describing the phase difference $\varphi (y,t)$ in the form

\begin{equation}
{1\over\omega_J^2}\,{\partial^2\varphi\over\partial t^2}+
\sin\varphi=l_J\int\limits_{-\infty}^\infty
dy'\,K\Bigl({y-y'\over 2\lambda_{\rm eff}}\Bigr)\,
\varphi''(y'),
\label{be25}
\end{equation}
where
\begin{equation}
K(u)={1\over\pi}\,
\int\limits_0^\infty {J_0(v)\over v+|u|}\,dv,
\label{be26}
\end{equation}
$J_0(v)$ is the zero-order Bessel function, and 
\begin{equation}
l_J={c\Phi_0\over 16\pi^2\lambda^2j_c}.
\label{be27}
\end{equation}

Note that Eq.~(\ref{be25}) can be rewritten as 
\cite{wephysica}
\begin{equation}
{1\over\omega_J^2}\,{\partial^2\varphi\over\partial t^2}+
\sin\varphi={l_J\over\pi}\,\int\limits_{-\infty}^\infty\,
{dy'\over y'-y}\,{\partial\varphi\over\partial y'}
\label{be28}
\end{equation}
in the limiting case when the characteristic space scale of
the phase difference variation is much less than 
$\lambda_{\rm eff}$. 

\section{Dispersion relation}
Let us now consider a small amplitude electromagnetic wave 
propagating along the Josephson junction. The corresponding
solution of Eq.~(\ref{be25}) then reads
\begin{equation}
\varphi=\varphi_0\,\exp (iky-i\omega t), \qquad 
|\varphi_0|\ll 1.
\label{de1}
\end{equation} 

Substitution of Eq.~(\ref{de1}) into Eq.~(\ref{be25})
results in the following dispersion relation
\begin{equation}
\omega=\omega_J\,\sqrt{1+
2k^2\lambda_{\rm eff}l_J{\cal K}(2k\lambda_{\rm eff})},
\label{de2}
\end{equation}
where 
\begin{equation}
{\cal K}(x)=\int\limits^{\infty}_{-\infty} 
K(u)\exp(ixu)\,du. 
\label{de2a} 
\end{equation}
The function ${\cal K}(x)$ has the following explicit form  
\begin{equation}
{\cal K}(x)=\cases{\displaystyle{1\over\pi\sqrt{1-x^2}}\,
\ln{{1+\sqrt{1-x^2}}\over
1-\sqrt{1-x^2}},
&if $x<1$;\cr
\ \ \ \cr
\displaystyle{1\over\sqrt{x^2-1}}\,
\Bigl[1-{2\over\pi}\,\arctan{1\over\sqrt{x^2-1}}\Bigr],
&if $x>1$.\cr} 
\label{de3}
\end{equation}

Using Eqs.~(\ref{de2}) and (\ref{de3}) we find, in 
particular, the dispersion relation $\omega (k)$ in the  
limiting cases $k\lambda_{\rm eff}\ll 1$ and
$k\lambda_{\rm eff}\gg 1$
\begin{equation}
\omega =\cases{\displaystyle\omega_J\sqrt{
1-{4k^2\lambda_{\rm eff}l_J\over\pi}
\ln(k\lambda_{\rm eff})},
&if\ \ $k\lambda_{\rm eff}\ll 1$;\cr
\ \ \ \cr
\omega_J\sqrt{1+kl_J},&if\ \ $k\lambda_{\rm eff}\gg 1$.\cr}
\label{de4}
\end{equation}

Thus, for $k\lambda_{\rm eff}\gg 1$ and $kl_J\gg 1$ the 
frequency of an electromagnetic wave propagating along the 
SIS-type tunnel Josephson junction in a thin film is 
proportional to the square root of the wave vector. 

The dispersion relation $\omega\propto\sqrt{k}$
leads, in particular, to a non-equidistant set of the
self-induced Fiske resonances in the current-voltage curves 
for the voltages $V>V_c$, where  
\begin{equation}
V_c={\hbar\omega(\lambda_{\rm eff}^{-1})\over 2e} 
\approx {c_s\over\lambda}\,\sqrt{d\over\lambda}.
\label{de5} 
\end{equation}

Note, that for a thin superconducting film the relation 
$k\lambda_{\rm eff}\sim 1$ corresponds to a wave length 
that is $\lambda /d$ times bigger than the London 
penetration depth.

\section{Summary}
To summarize, we have found the integro-differential 
equation describing the phase difference in case of 
the SIS-type tunnel Josephson junction in a thin 
superconducting film. We apply this equation to calculate 
the dispersion relation for an electromagnetic wave 
propagating along the Josephson contact. We have shown 
that if the wave length is small compared with the Pearl 
penetration depth $\lambda_{\rm eff}$ the frequency is 
proportional to the square root of the wave vector.

\acknowledgments
We are grateful to Dr.~E.~Polturak for useful discussions.
This work was supported in part by the Foundation Raschi.

\begin{figure}
\caption{A thin superconducting film with a Josephson 
junction (thick line).}
\label{f1}
\end{figure}

\end{document}